\documentclass{aip-cp}

\usepackage[numbers,square,sort&compress]{natbib}
\usepackage{txfonts}
\usepackage{graphicx}
\graphicspath{{./graphics/}}

\title{Polarization Observables using Positron Beams}

\author[mit]{Axel Schmidt}
\eaddress{schmidta@mit.edu}

\affil[mit]{Massachusetts Institute of Technology, Cambridge, MA, USA}

\begin{document}

\maketitle

\begin{abstract}
The discrepancy between polarized and unpolarized measurements of the proton's electromagnetic
form factors is striking, and suggests that two-photon exchange (TPE) may be playing a larger
role in elastic electron-proton scattering than is estimated in standard radiative corrections
formulae. While TPE is difficult to calculate in a model-independent way, it can be determined
experimentally from asymmetries between electron-proton and positron-proton scattering. The
possibility of a polarized positron beam at Jefferson Lab would open the door to measurements
of TPE using polarization observables. In these proceedings, I examine the feasibility of
measuring three such observables with positron scattering. Polarization-transfer, specifically
the $\epsilon$-dependence for fixed $Q^2$, is an excellent test of TPE, and the ability to compare
electrons and positrons would lead to a drastic reduction of systematics. However, such a measurement
would be severely statistically limited. Normal single-spin asymmetries (SSAs) probe the imaginary
part of the TPE amplitude and can be improved by simultaneous measurements with electron and positron
beams. Beam-normal SSAs are too small to be measured with the proposed polarized positron beam,
but target-normal SSAs could be feasibly measured with unpolarized positrons in the spectrometer
halls. This technique should be included in the physics case for developing a positron source
for Jefferson Lab.
\end{abstract}

\section{Introduction}

The discrepancy between the proton's electromagnetic form factor ratio, $\mu_p G_E / G_M$, 
extracted from polarization asymmetry measurements and the ratio extracted from unpolarized
cross section measurements leaves the field of form factor physics in an uncomfortable state
(See \cite{Afanasev:2017gsk} for a recent review).
On the one hand, there is a consistent and viable hypothesis that the discrepancy is caused
by non-negligible hard two-photon exchange (TPE) \cite{Guichon:2003qm,Blunden:2003sp}, the one radiative correction omitted from 
the standard radiative correction prescriptions \cite{Mo:1968cg,Maximon:2000hm}. On the 
other hand, three recent measurements 
of hard TPE---at VEPP-3, at CLAS, and with OLYMPUS---found that the effect of TPE is small
in the region of $Q^2 < 2$~GeV$^2/c^2$ \cite{Rachek:2014fam,Adikaram:2014ykv,Rimal:2016toz,Henderson:2016dea}. 
The TPE hypothesis is still viable; it is possible 
that hard TPE contributes more substantially at higher momentum transfers. However, the three
recent TPE experiments were challenging measurements, and the idea that a subsequent suite of 
follow-on measurements could quickly map out hard TPE over a large kinematic space with small uncertainty
does not seem realistic.

In the face of this challenge, it may be more productive to concentrate experimental effort
on constraining and validating model-dependent theoretical calculations of TPE. There are 
multiple theoretical approaches, with different assumptions and different regimes of validity
\cite{Chen:2004tw,Afanasev:2005mp,Tomalak:2014sva,Blunden:2017nby,Kuraev:2007dn}.
If new experimental data could validate and solidify confidence in one or more theoretical
approaches, then hard TPE could be treated in the future like any of the other standard radiative
corrections, i.e., a correction that is calculated, applied, and trusted. 

VEPP-3, CLAS, and OLYMPUS all looked for hard TPE through measurements of the positron-proton
to electron-proton elastic scattering cross section ratio. After applying radiative corrections,
any deviation in this ratio from unity indicates a contribution from hard TPE. However, this is
not the only experimental signature one could use. Hard TPE can also appear in a number of polarization
asymmetries. Having constraints from many orthogonal directions, i.e., from both cross section 
ratios and various polarization asymmetries would be valuable for testing and validating theories
of hard TPE. As with unpolarized cross sections, seeing an opposite effect for electrons and positrons
is a clear signature of TPE. 

In the following sections, I examine the feasibility of measuring three different types of polarization
asymmetries with a hypothetical future positron beam at Jefferson Lab. I find that polarization 
transfer would be a systematically clean technique, but would be hampered by poor statistical precision.
Beam normal single spin asymmetries would not be feasible due to the high luminosities required. 
Target normal single spin asymmetries are feasible from the perspective of statistics; the challenge
would then become finding ways to reduce systematic uncertainty.

\section{Polarization Transfer}

The polarization transfer technique is the most accurate way of measuring the proton's form
factor ratio. A polarized electron beam is scattered from an unpolarized proton target, and
the polarization of the recoiling proton is then measured using a secondary scattering reaction.
The ratio of transverse polarization to longitudinal polarization (within the scattering plane)
is proportional to the form factor ratio. This technique has been employed in experiments 
covering a wide range of momentum transfers, including those at MIT Bates \cite{Milbrath:1997de}, Mainz \cite{Pospischil:2001pp}, 
and Jefferson Lab \cite{Gayou:2001qt, MacLachlan:2006vw,Ron:2011rd,Paolone:2010qc,Zhan:2011ji},
including three experiments, GEp-I \cite{Punjabi:2005wq}, GEp-II \cite{Punjabi:2005wq}, and GEp-III \cite{Puckett:2017flj} 
that pushed to high momentum transfer. Another experiment, GEp-2$\gamma$, looked for hints of 
TPE in the $\epsilon$-dependence in polarization transfer \cite{Puckett:2017flj}. Two other experiments made equivalent
measurements by polarizing the proton target instead of measuring recoil polarization \cite{Jones:2006kf,Crawford:2006rz}.

While polarization transfer is less sensitive to the effects of hard TPE, it is not immune. 
Following the formalism of Ref. \cite{Carlson:2007sp}, one finds that
\begin{equation}
\frac{P_t}{P_l} = \sqrt{\frac{2\epsilon}{\tau (1+\epsilon)}}\frac{G_E}{G_M} \times \left[ 1+ 
  \textrm{Re}\left(\frac{\delta \tilde{G}_M}{G_M}\right) + \frac{1}{G_E}\textrm{Re}\left(\delta\tilde{G}_E + \frac{\nu}{M^2}\tilde{F}_3\right)
    - \frac{2}{G_M}\textrm{Re}\left(\delta\tilde{G}_M + \frac{\epsilon\nu}{(1+\epsilon)M^2} \tilde{F}_3\right) + \mathcal{O}(\alpha^4) \right],
\end{equation}
where $M$ is the mass of the proton, $\tau\equiv Q^2/4M^2$, $\epsilon^{-1} \equiv 1 + 2(1+\tau)\tan^2\frac{\theta}{2}$, 
$\nu \equiv (p_e + p_{e'})_\mu (p_p + p_{p'})^\mu$, and where $\delta \tilde{G}_E$,  $\delta \tilde{G}_M$, and
$\delta \tilde{F}_3$ are additional form factors that become non-zero when moving beyond the one-photon exchange
approximation. This particular dependence on new form factors is slightly different than one what finds when
taking a positron to electron cross section ratio:
\begin{equation}
\frac{\sigma_{e^+p}}{\sigma_{e^-p}} = 1 + 4G_M\textrm{Re}\left(\delta \tilde{G}_M + \frac{\epsilon \nu}{M^2}\tilde{F}_3\right)
- \frac{4\epsilon}{\tau}G_E\textrm{Re}\left(\delta \tilde{G}_E + \frac{\nu}{M^2} \tilde{F}_3\right) + \mathcal{O}(\alpha^4).
\end{equation}
A measurement of the difference in polarization transfer between electron and positron scattering therefore adds
information about TPE in addition to what can be learned from cross section ratios alone.

The GEp-2$\gamma$ experiment looked for the effects of TPE in polarization transfer by making measurements
at three kinematic points with varying values of $\epsilon$, but with $Q^2$ fixed at 2.5~GeV$^2/c^2$ \cite{Puckett:2017flj}.
Since in the absence of hard TPE the ratio $G_E/G_M$ has no $\epsilon$-dependence, any variation with $\epsilon$
is a sign of hard TPE. The GEp-2$\gamma$ measurement was statistically consistent with no $\epsilon$-dependence,
though their measurement of purely the longitudinal component showed deviations from the one-photon exchange
expectation.

Positron scattering has the potential to improve the capabilities of a polarization transfer measurement. The largest
systematic uncertainties, spin precession in the magnetic spectrometer and polarimeter alignment, are associated with 
proton polarimetry. These effects will largely cancel when taking a super ratio $(P_t/P_l)_{e^+p} / (P_t/P_l)_{e^-p}$,
producing a systematically clean measurement. Ideally one would measure polarization transfer in both beam species for 
points at fixed $Q^2$ and look for $\epsilon$-dependence as a signature. The figure-of-merit for such a measurement
would be 
\begin{equation}
\label{eq:pt_fom}
\textrm{F.o.m.} \sim A P_e \sqrt{\frac{\textrm{d}\sigma}{\textrm{d}\Omega} \Omega\mathcal{L} T \varepsilon },
\end{equation}
where $A$ is the spectrometer analyzing power, $P_e$ is the lepton beam polarization, d$\sigma/$d$\Omega$ is the
elastic cross section, $\Omega$ is the spectrometer acceptance, $\mathcal{L}$ is the luminosity, $T$ is the run
time, and $\varepsilon$ is the running efficiency, i.e., the ratio of the live-time to wall-time.

I have attempted to gauge the reach of such an experiment with a hypothetical future positron beam 
at Jefferson Lab. I envisioned an experiment in Hall C, along the same lines as GEp-III and GEp-2$\gamma$,
but using both the HMS and SHMS spectrometers to detect recoiling protons. Both spectrometers would need to be equipped
with focal plane polarimeters. As in GEp-III and GEp-2$\gamma$, a non-magnetic, calorimetric detector,
such as BigCal, could be used to detect scattered leptons in coincidence. I envisioned using two BigCals, each
paired to match the acceptance of one of the spectrometers. Rather than calculate the statistical precision
directly, I have scaled the statistical uncertainty from GEp-III and GEp-2$\gamma$ using the figure-of-merit
from equation \ref{eq:pt_fom}, i.e., matching the analyzing power and running efficiency from those experiments.
There are two parameters in the figure-of-merit that would have significantly different values from the 
previous experiments. The achievable positron polarization would be $\approx 60\%$, down from the $\approx 80\%$ polarization 
of the current CEBAF beam. The positron beam would also be limited to 100~nA. GEp-III and GEp-2$\gamma$ used
80~$\mu$A. These two effects mean that a future positron scattering experiment would have a factor 38 worse
statistical uncertainty before run time and cross section are taken into account.

\begin{figure}[htpb]
\centering
\includegraphics[width=0.5\textwidth]{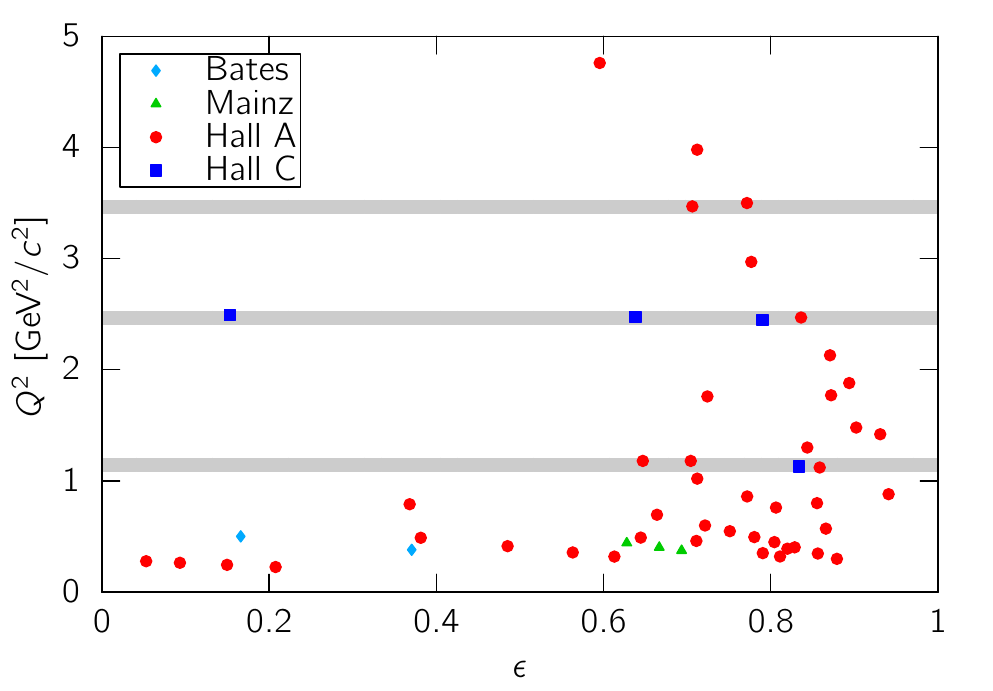}
\caption{\label{fig:pt_kin} $Q^2=1.15$, 2.5, and 3.5~GeV$^2/c^2$ already have multiple prior
polarization transfer measurements, and would therefore be sensible targets for an experiment
with positrons.
}
\end{figure}

It would be sensible to measure at values of $Q^2$ where there have already been prior polarization transfer
experiments with which to compare. The values of $\epsilon$ and $Q^2$ for all previous polarzation transfer
measurements are shown in figure \ref{fig:pt_kin}. Because of prior measurements at these values, I initially 
selected $Q^2=1.15$, 2.5, and 3.5~GeV$^2/c^2$ as candidate kinematics. The experiment would need to cover 
several different $\epsilon$ points. The SHMS, which is limited to more forward scattering angles, could stay 
fixed for entire run to detect protons at $\epsilon=0.2$, while the HMS could move between $\epsilon=0.5$ and 
0.8, where cross sections are higher and statistics can be accumulated more quickly. 

\begin{figure}[htpb]
\centering
\includegraphics[width=0.5\textwidth]{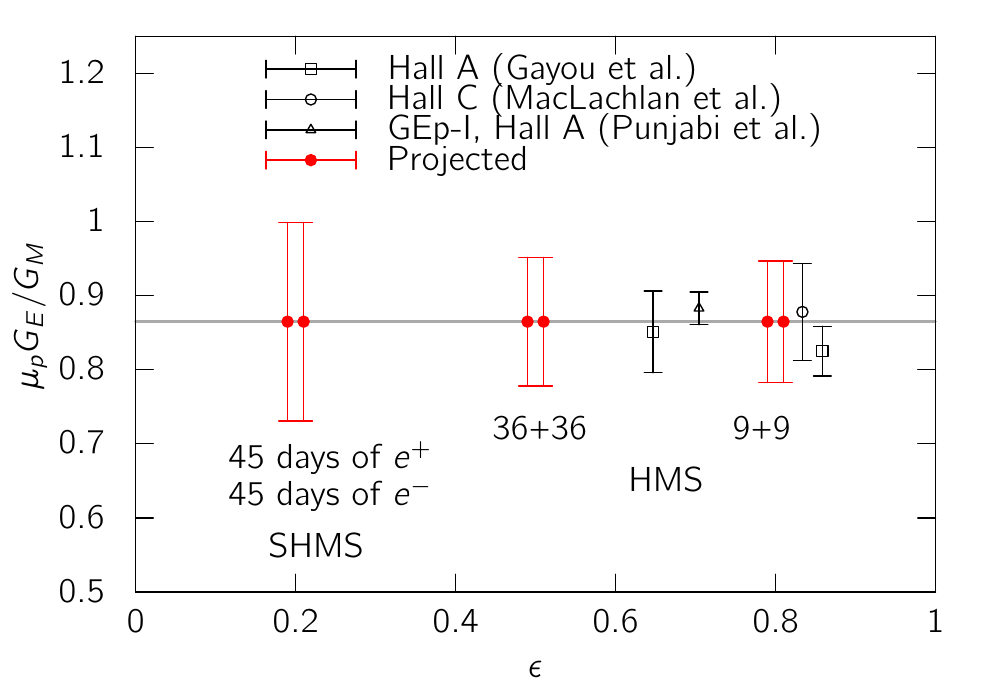}
\includegraphics[width=0.5\textwidth]{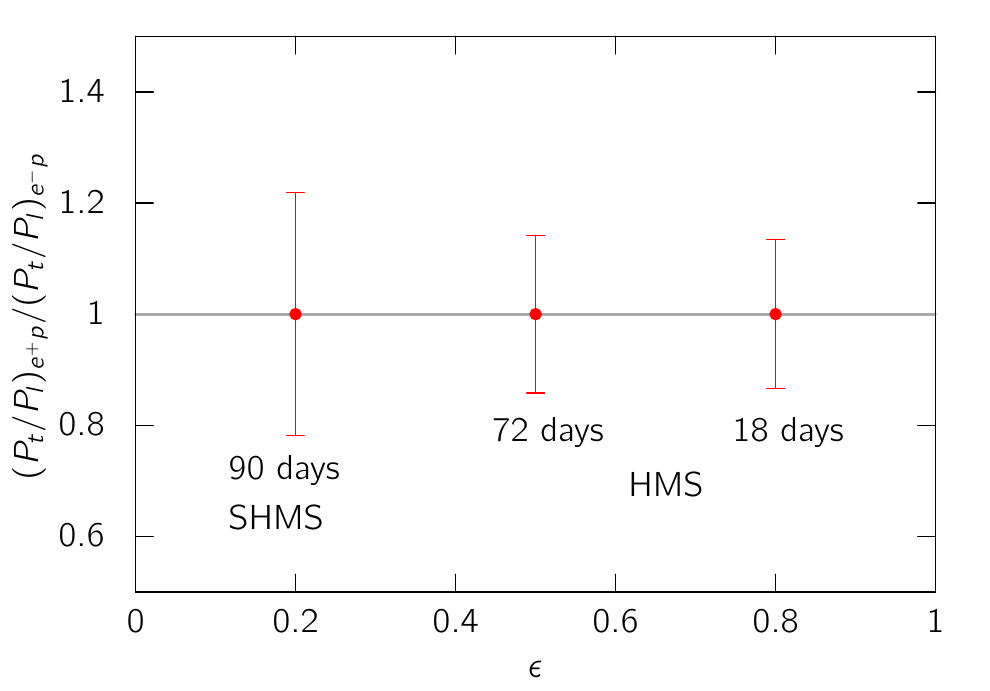} 
\caption{\label{fig:pt_reach} Even with a 90-day measurement at a limited $Q^2$ of only 1.15~GeV$^2/c^2$, 
a polarization transfer measurement with positrons would be limited to 10--20\% statistical precision, 
hardly enough to constrain a percent level TPE effect.}
\end{figure}

The GEp-2$\gamma$ experiment collected data for 40 days. The projected uncertainty for a 90-day experiment
with both electrons and positrons at $Q^2=1.15$~GeV$^2/c^2$ is shown in figure \ref{fig:pt_reach}. Even in
90 days and at reduced $Q^2$, the projected uncertainties are on the order of 10\% on $\mu_p G_E/G_M$ and
10--20\% on $(P_t/P_l)_{e^+p} / (P_t/P_l)_{e^-p}$. This would be insufficient for constraining a percent-level
TPE effect. For larger $Q^2$, where the elastic cross section is considerably lower, these projections become
even more pessimistic. Despite the polarization transfer ratio between positrons and electrons being a
systematically clean technique, it would not be feasible statistically, given the 
design parameters for a future positron beam at Jefferson Lab. 

A measurement of the ratio of tranverse to longitudinal polarization transfers is equivalent to a measurement
of the ratio of transverse to longitudinal beam-target double spin asymmetries, where the longitudinal and transverse
directions are relative to the momentum transfer vector. Because of this equivalence, one could imagine taking advantage
of the large acceptance of the CLAS12 spectrometer and performing such a measurement in Hall B. Unfortunately,
this technique relies on having some component of the target polarization being transverse to the beam direction, so that
the double spin asymmetry manifests as an azimuthal asymmetry for constant scattering angle. As of now, the CLAS12
solenoid, which has a 5~T field parallel to the beam direction, prevents the use of any non-longitudinal target
polarizations, making this technique unfeasible for now.

\section{Beam-Normal Single Spin Asymmetries}

Single spin symmetries (SSAs) normal to the reaction plane are zero in the limit of one-photon exchange, and
thus make an excellent signature for detecting two-photon exchange. There are two such asymmetries: beam-normal,
in which a transversely polarized lepton beam is scattered from an unpolarized target, and target-normal, 
in which an unpolarized lepton beam is scattered from a transversely polarized target. Both of these asymmetries
are sensitive to the imaginary parts of the additional multi-photon exchange form factors. The imaginary parts
do not contribute at $\alpha^3$ order to the elastic cross section, meaning that they do not play any role in
the proton form factor discrepancy. However, they can provide valuable orthogonal constraints to theories of
TPE. 

\begin{figure}[htpb]
\centering
\includegraphics[width=0.5\textwidth]{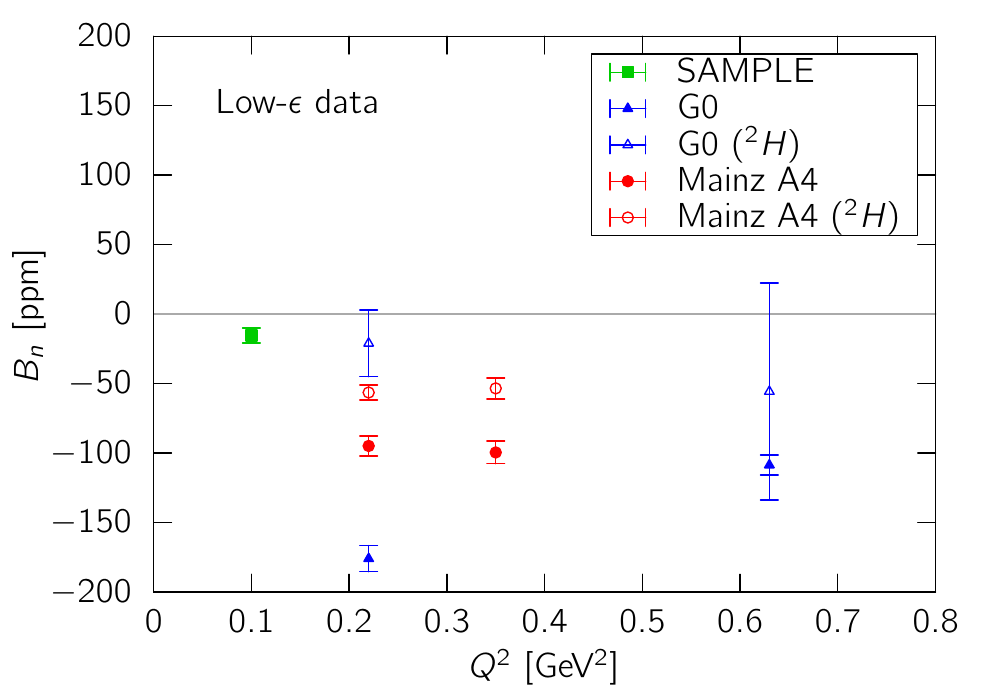}
\includegraphics[width=0.5\textwidth]{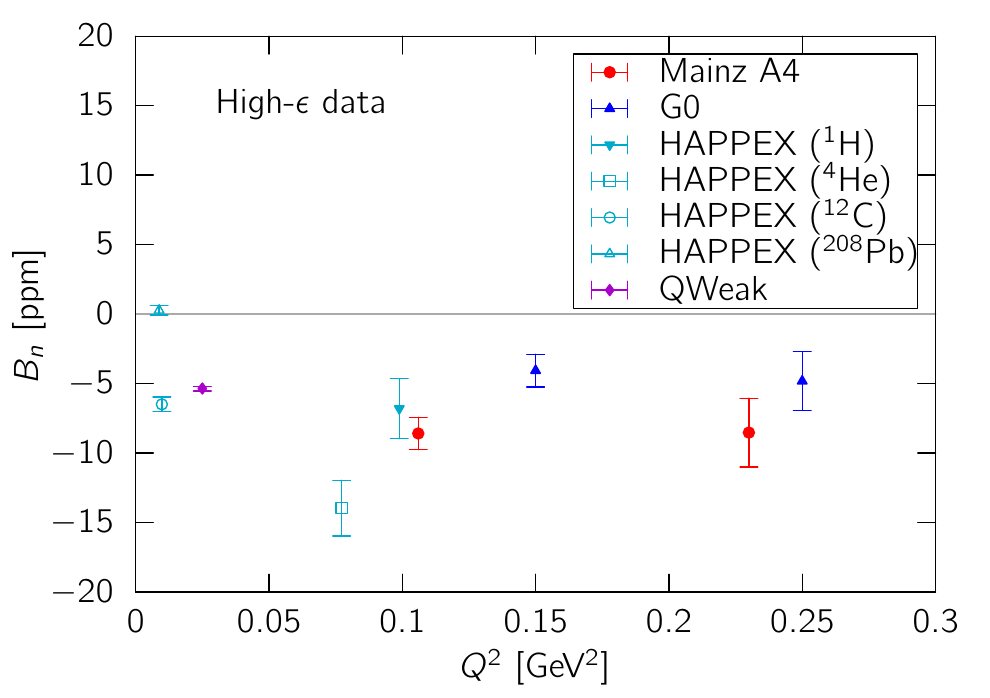} 
\caption{\label{fig:bn_data} Previous $B_n$ measurements at low-$\epsilon$ (left panel) have found asymmetries on
the order of 100~ppm, while at high-$\epsilon$ (right panel), the asymmetries are even smaller. 
}
\end{figure}

The beam-normal SSA, $B_n$, depends on the additional form factors in the following way:
\begin{equation}
B_n = \frac{4mM\sqrt{2\epsilon(1-\epsilon)(1+\tau)}}{Q^2(G_M^2 + \frac{\epsilon}{\tau}G_E^2)} \times
\left[ -\tau G_M \textrm{Im}\left( \tilde{F}_3 + \frac{\nu}{M^2(1+\tau)}\tilde{F}_5\right)
  -G_E\textrm{Im}\left(\tilde{F}_4 + \frac{\nu}{M^2(1+\tau)}\tilde{F}_5\right)\right] + \mathcal{O}(\alpha^4),
\end{equation}
where $m$ represents the electron mass, and $\tilde{F}_4$ and $\tilde{F}_5$ are yet additional
new form factors. Beam-normal effects are helicity suppressed, hence the leading factor of the electron
mass, and, as a result, beam-normal SSAs are approximately three orders of magnitude smaller than
target-normal SSAs, on the order of 10--100 ppm. Nevertheless, beam normal SSAs have been measured in 
electron scattering by several experiments and have been shown conclusively to be non-zero. Beam-normal 
SSAs are a possible source of false asymmetry in parity-violating electron scattering (PVES) experiments, and
because of this, nearly every major PVES experiment has produced beam-normal SSA measurement as well,
including, but not limited to, SAMPLE \cite{Wells:2000rx}, G0 \cite{Armstrong:2007vm, Androic:2011rh}, 
A4 \cite{Maas:2004pd,Rios:2017vsw}, HAPPEX/PREX \cite{Kaufman:2007zz,Abrahamyan:2012cg}, and 
Q-Weak \cite{Waidyawansa:2016znm}. The data from previous $B_n$ measurements are shown in figure \ref{fig:bn_data}.

While a 100~ppm asymmetry is far easier to measure than a 0.1~ppm parity-violating asymmetry, a measurement
with a 100~nA polarized positron beam at Jefferson Lab would not be practical. PVES experiments typical use 
beam currents with many tens if not hundreds of $\mu$A. Furthermore, a positron beam does not provide any
benefit in reducing the main systematic uncertainties: uncertainty in beam polarimetry and false asymmetries
produced by the detectors.

\section{Target-Normal Single Spin Asymmetries}

The target-normal SSA, $A_n$, depends on the additional form factors in the following way:
\begin{equation}
A_n = \frac{\sqrt{2\epsilon(1+\epsilon)}}{\sqrt{\tau}\left(G_M^2 + \frac{\epsilon}{\tau}G_E^2\right)} \times
\left[ -G_M \textrm{Im}\left( \tilde{G}_E + \frac{\nu}{M^2}\tilde{F}_3 \right) 
+ G_E \textrm{Im} \left( \tilde{G}_M + \frac{2 \epsilon \nu}{M^2 (1+\epsilon)} \tilde{F}_3 \right) \right] + \mathcal{O}(\alpha^4).
\end{equation}
Since this is also a unique combination of new form factors, a measurement of $A_n$ would
provide new constraints on TPE. Previous measurements of $A_n$ with electron scattering 
have either been made with inelastic scattering \cite{Chen:1968mm,Appel:1970ni,Rock:1970sj,Airapetian:2009ab,Katich:2013atq}, 
or in quasielastic scattering from polarized $^{3}$He \cite{Zhang:2015kna}. There are 
currently no published results from elastic electron scattering from polarized hydrogen. 
Ref. \cite{Zhang:2015kna} measured an asymmetry of a few parts per thousand in $^{3}$He,
which corresponds to an asymmetry of a few percent from polarized neutrons. It would be
reasonable to expect an asymmetry of similar size from polarized protons. 

Since $A_n$ is zero in the one-photon exchange approximation, any measurement of non-zero
$A_n$, either in positron scattering or in electron scattering, would reveal either TPE,
or a contribution from a T-violating process. As an asymmetry from TPE will flip sign when
switching between electrons and positrons, a measurement with both electrons and positrons
can distinguish between T-violation and the effects of TPE.
Unfortunately, unlike in the case of polarization transfer, positrons do not offer any
avenues for reducing the systematic uncertainties, which, for a measurement
of $A_n$, are associated with the target polarization (including dilution factors), 
target alignment, live-time asymmetry, and radiative corrections. 

\begin{figure}[htpb]
\centering
\includegraphics[width=0.6\textwidth]{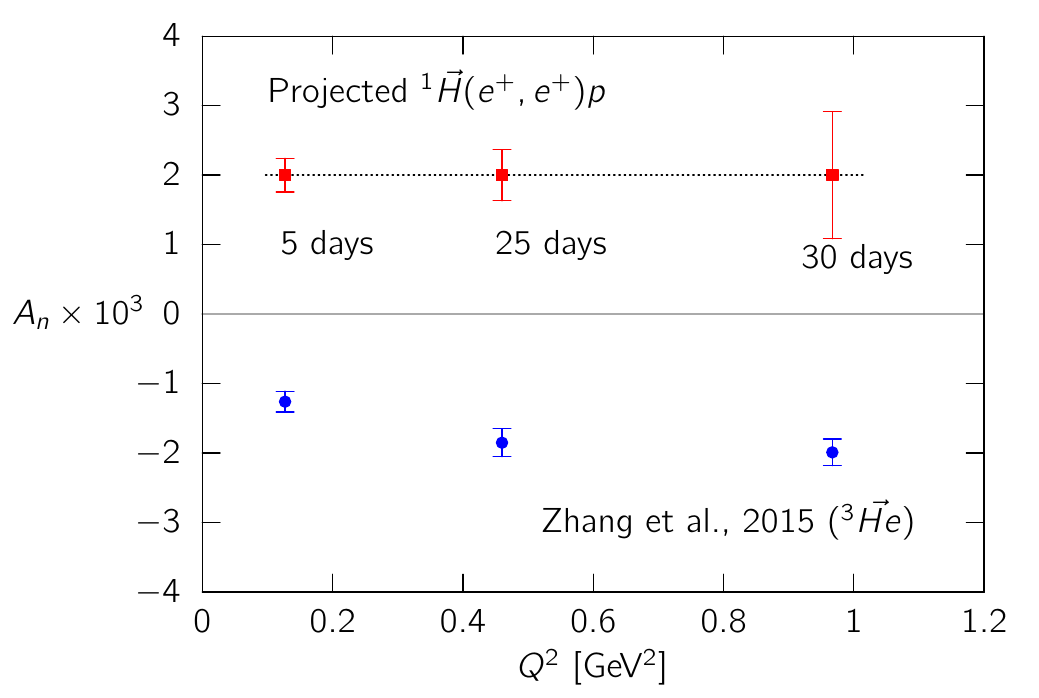}
\caption{\label{fig:an_reach} A measurement of $A_n$ using positrons on a polarized proton
target would be a feasible from the perspective of statistics. However, the challenge remains to successfully 
rein in systmatic uncertainties.}
\end{figure}

I have attempted to gauge the feasibility of measurement of $A_n$ in elastic scattering from
transversely polarized protons using a hypothetical CEBAF positron beam. The highest achievable luminosity
with a polarized proton target would be attained using a frozen ammonia target, such as the one 
developed for the $g_2^p$ and $G_E^p$ experiments \cite{Pierce:2014xma}. Such a target would be limited to 100~nA of current, but would
allow running with $\mathcal{L} = 10^{35}$~cm$^{-2}$s$^{-1}$, with a dilution factor of 12\%. 
I assumed an experimental set-up along the same lines as ref. \cite{Zhang:2015kna}, with both Hall A
HRS spectrometers used simultaneously to double the acceptance for scattered positrons. 
The projected statistical uncertainty is shown in figure \ref{fig:an_reach}, assuming a two
month measurement with 50\% live-time, with data points for 1.2, 2.4, and 3.6~GeV beams and 
both HRSs positioned at $17^\circ$. The projections indicate that such a measurement could attain
adequate statistics. The challenge would remain to control systematic uncertainties. As the 
frozen spin target polarization decays due to irradiation by the beam, frequent and accurate polarimetric
measurements would be critical. If the measurement were combined with an electron scattering
measurement (which would not be essential), frequent flipping between $e^-$ and $e^+$ modes would
be desired.

\section{Conclusions}

Polarization asymmetries in positron scattering would provide valuable new information that can test theories
of hard TPE. I have considered three possible asymmetries and assessed the feasibility of measurements in the
spectrometer halls with a future positron beam at Jefferson Lab. In the first category of polarization transfer
measurements, positrons can dramatically improve the systematic uncertainties. However, with a polarized positron
beam current of 100~nA, only 10--20\% measurements would be possible, due to limited statistics. The second category,
beam-normal single spin asymmetries, an enormous luminosity is needed to resolve the 1--100 ppm asymmetry. The third
category, target-normal single spin asymmetry, holds some promise. Though positrons do not offer any avenue to
reducing systematic uncertainty, such a measurement would be feasible from the perspective of statistics with a 100~nA beam and a
frozen ammonia target. By comparing asymmetries in positron- and electron-scattering, the effects of TPE could
be definitely identified, and contributions from possible T-violation could be ruled out. This avenue deserves
further study and should be included as part of the physics case for building a positron source at CEBAF. 

\section{Acknowledgements}

This work was supported by the Office of Nuclear Physics of the U.S. Department of Energy, grant No. DE-FG02-94ER40818.

\bibliographystyle{aipnum-cp}
\bibliography{references}

\end{document}